\input amstex

\magnification=\magstep 1 \baselineskip=24pt 
\vskip .5in 
 
 \centerline{\bf      The Problem of Quantum Measurement} 
\centerline {by Joseph F.\ Johnson} 
\baselineskip=12pt  
\centerline {Math Dept., Univ.\ of New Hampshire} 
\baselineskip=24pt 
\vskip .3in 
 
\centerline{\bf Abstract} \baselineskip=8pt  
\font\eightrm=cmr8 
\eightrm 
 
We derive the probabilities of measurement results from Schroedinger's  
equation plus a definition of macroscopic as a particular kind of thermodynamic limit.  
Bohr's insight that a measurement apparatus must be classical in nature and classically 
describable is made precise in a mathematical sense analogous to the procedures of classical 
statistical mechanics and the study of Hamiltonian heat baths.   
 
\baselineskip=24pt  
\rm 
 
 
 

          
\centerline     {\bf Quantum Measurement as Thermodynamic Limit}

It is not necessary to modify the axioms of non-relativistic Quantum Mechanics in order
to solve the puzzle of Quantum Measurement.  In order to do so, all we need to do is to 
take Schroedinger's equation without any modifications as basic, imitate the procedures
of classical Statistical Mechanics, and use an explicit Hamiltonian dynamics that models
the amplification process.  No measurement apparatus dispenses with some form of 
amplification, and it should not be a surprise that the key to the problem is to use the 
physics of real measurement apparatuses.  

    \centerline {Introductory remarks}

     Quantum Mechanics has various axioms, say about six or so.  Three of them are logically of the same structure as the formulation of Hamiltonian mechanics, and are determinstic.  Famously, the time-evolution of a quantum system, as long as it is unobserved, is governed by the wave equation of Schroedinger, which is in its mathematical structure, a deterministic equation.  It describes a deterministic dynamics.  But during the measurement process, a stochastic dynamics supervenes, and only probability laws describe the result of a measurement and the state of a particle as it escapes the measurement apparatus.     The problem of Quantum Measurement can be roughly stated as does Bell in a famous paper, ``Against Measurement,'' as well as other papers, where do we put the cut in Nature that describes when one dynamical law rules, and when the other law rules?

     There have been, perhaps, three main obstacles to solving the problem.  Only one has 
been purely physical.  That one is the lack of writing down an explicit, time-independent, 
matrix Hamiltonian which, even in a toy model, captures the properties of the amplification process which are physically relevant to measurement apparatuses.  This is 
achieved for perhaps the first time in this paper.  The other two are merely formal: as 
experts in the philosophy of science know, there has been heretofore no widely accepted 
definition of the concept of ``probability'' as it occurs in science, and therefore as it occurs in the axioms of quantum measurement.  We adopt one very close to Jan von Plato's, and it is a powerful argument in its favour that it, originally advanced purely in the context of classical mechanics, works just as well in this new setting.  It is close to the usual working but inadequate frequency theory of probability, which again should be no surprise, since the frequency theory of probability has worked well in science in spite of its logical circularity (a circularity fixed by von Plato's adjustments).  

     The last obstacle has been common misconceptions about the logical structure of 
classical Statistical Mechanics, but fortunately there are in some standard texts and 
standard and influential papers an adequate discussion of this so we need only imitate 
the procedure of, for example, the celebrated paper of Ford--Kac--Mazur.  Each of these philosophical points will also be treated in detail in the appropriate section of this paper.

     A more careful examination of the logic of Quantum Mechanics and how the idea of measurement could be rigourously defined within it shows, as we will see, that the crucial missing ingredient is a rigourous definition of ``macroscopic.''  This seems not to have been even clearly noticed before.  Imagine, as usual, that the amplifying apparatus doing the measuring registers the result of the measurement by having a needle move along a dial, pointing in a macroscopically visible fashion to different numbers.  
If we formulate our task, as it has often been formulated, to be that of deriving the quantum-mechanical measurement postulates 
from Schroedinger's equation as applied to the joint system of microscopic particle 
being measured, and amplifying apparatus measuring it, it has always been said that 
the apparatus has various possible states corresponding to different positions of, say,
 its needle in the gauge.  These visibly different numbers or positions of the needle 
are called, ``pointer positions.''  The dynamic variable of the apparatus 
that discriminates between different pointer positions is called a (or the) ``pointer variable.''  It seems not to have been realised 
that it was a wide open question what sort of mathematical object should model such pointer variables, or what was the physical basis for making assumptions about their properties.  Our view is that no ad hoc assumptions are allowable, we must give a rigourous definition of pointer variable in terms of the basic notions of quantum 
mechanics and derive from Schroedinger's equation alone whatever properties of
 pointer variable we wish to rely on---always in a way according with experimental 
results and physical intuition, of course.  This is the main, or even only, novelty 
in this paper.  Every other idea has appeared in print before, only not combined with 
the other ones, or even unfortunately in combination with usuitable
ideas on the other ingredients.

Einstein had already suggested that the probabilities of quantum mechanics arose from 
a fundamental deterministic dynamics in analogy with the way the probabilites of 
classical statistical mechanics arose from Newtonian mechanics.  Bohr already suggested 
that measurement apparatuses and pointer positions were classical objects, not quantum, and classically describable.  Schroedinger had already suggested that Schroedinger's equation would not need to be replaced by a different deterministic dynamics.  Daneri--Prosperi--Loinger had already suggested that amplification by a macroscopic device in a meta-stable state was physically key to measurement, and an ergodic principle of some sort was at work.  H.S. Green had already suggested that a device in a state of negative temperature was an appropriate model for the measurement apparatus.  Schwinger had already suggested that a negative temperature Brownian motion would amplify quantum motion to a macroscopic level of classical motion in which the quantum uncertainties would be negligible (although this was not in the context of quantum measurement).

But, for example, Einstein's idea may have been irrelevantly tied up with the idea that hidden variables were essential to classical statistical mechanics, an idea later refuted by 
Darwin and Fowler.  Green's treatement of measurement involved postulating a probability distribution on the measurement apparatus, which is an unduly na\"ive way to ground Statistical Mechanics.  The Coleman--Hepp model of measurement 
both lacks the notion of amplification and imports the techniques of Quantum Statistical Mechanics.  (The ideas of Quantum Statistical Mechanics are physically inappropriate to 
the problem of Quantum Measurement if Bohr's insight that measurement apparati are classical in nature
\ is correct.  Because the thermodynamic limit taken in Quantum Statistical Mechanics is 
still a quantum system, not a classical system.)  
Bohr's suggestion was usually phrased in terms of accepting a dualism in physics or 
even the `cut' in Nature Bell complained about.  He failed to connect it with a 
thermodynamic limit, any definite dynamical content, or a precisely defined approximation 
procedure.  He once more carefully specified what he meant by saying that the measurement 
apparatus in classical.  He meant that it could be usefully described in an approximation 
in which Planck's constant was neglected.

\centerline {Precise statement of the problem}

It is helpful to focus on Wigner's formulation of the problem, rather than Bell's.  To do this, we state the six axioms of Quantum Mechanics in Dirac's formulation--assuming, as usual, for simplicity, that only discrete eigenvalues with multiplicity one occur, etc.\ etc.  
The first three axioms are the same as those of Hamiltonian Mechanics, with only technical mathematical differences.  The first one is  that each closed physical system is described by a Hilbert space and a Hamiltonian operator on that space, which is characteristic of that system.  That is, to each system is associated a complex separable Hilbert space $\Cal H$ and a skew-adjoint operator $H$ defined on a dense subspace.  These mathematical objects give us all the physical information about that 
system.  
The second axiom is that the possible physical states of the system are given 
 by non-zero elements $\psi$ of $\Cal H$, called wave-functions,  
and if and only if $\psi_1=c\psi_2$ with $0\neq c \in \bold C$ do they
describe   the same physical state.   
 
The third axiom says that if the system is in the state $\psi_o$  
at time $t=0$ then it will be in the state  
$$\psi_t = e^{2\pi t\frac Hh}\cdot \psi_o$$ 
at time $t$, where $h$ is Planck's constant, roughly 9$\times$ $10^{-37}$ hp-sec$^2$. 
 
If Quantum Mechanics were really the same as Classical Mechanics, this would be  
all the axioms needed.  One would operationalise somehow the lab procedures needed  
to measure the various $\psi$ of various types of physical systems. 
But instead of this, Quantum Mechanics introduces two new undefined, primitive concepts  
into the axioms.  In Classical Mechanics, what one measured were the states,  
directly.  This is no longer true, one now introduces a new undefined concept, supplementing  
that of physical state.  One also makes measurement a specific concept, although primitive  
and undefined, it is specifically different from other physical processes. 
 
There are three axioms about the measurement process.  The first one is that to  
every possible measurement process, there corresponds a self-adjoint operator Q such that its  
eigenvectors, or eigenstates, form an orthonormal basis of $\Cal H$.  Let the  
orthonormal basis be $\{ \psi_i\}$  
and let the associated eigenvalues be $\lambda_i$ so that we have $Q\psi_i = \lambda 
_i \psi_i$.  The only possible results of the measurement process are the  
eigenvalues, $\lambda_i$.  The next axiom states that if the system being measured is  
in the state $\psi$ and if the Fourier decomposition of $\psi$ is given by   
$$\psi = \sum_i c_i \psi_i$$ 
then the probability that the result of the measurement process will be $\lambda_i$ is  
$\vert c_i \vert ^2$ if we assume, as we may after an inessential normalisation factor  
is inserted, that $\vert \vert \psi \vert \vert ^2 =1$ or, as we say, that $\psi$ is normalised. 
 
The last 
is the reduction of the wave packet: it states that after the measurement process is over,  
the system being measured is in the physical state $\psi_i$ corresponding to the eigenvalue  
$\lambda_i$ which was observed. 
 
(Dirac's original line of argument  
for it was based on reasoning using the principle of continuity, 
Most physicists have given up interpreting this axiom literally.  For this reason, we reserve discussion of 
this axiom for a projected sequel.) 
 
Although many discussions of the problem of Quantum Measurement have focussed on this last axiom, 
Wigner's influential discussion does not.  
Most discussions or questions that occur to the beginner about the problem of Quantum 
Measurement involve hidden assumptions, often of a philosophical nature, in addition 
to the axioms.  For example, by now it has been realised that although the process is 
called `measurement,' it must not be assumed that there is, physically, some `value' which is being measured in the sense that it is pre-existing.  


J.S.Bell has influentially intervened in this project several times.  For us, 
the two most relevant times are in his comments on the Coleman--Hepp 
model and, by extension, most statistical approaches.  And in one of his last 
articles, `Against Measurement' in which he critiques the standard approaches 
even more forcefully.

For us, his critiques come down to three points: The orthodox approach is not 
physically precise if it cannot give a physical definition of measurement apparatus 
in terms of wave functions and Hilbert Spaces and Hamiltonian operators.  (It 
should not be left to the experience and tact of the theoretician.)  The theory, in 
order to even *be* a theory, must also be capable of being logically precise: 
assumptions must be distinguished from theorems, and primitive undefined 
concepts must be carefully laid out at the start, with all further concepts defined 
in terms of them.  And the theory should be `about' reality, all of reality, including 
the universe as a whole with all measurement apparatuses included in it.

(In particular about being logically precise, Bell makes two points which we 
will answer: the first one he phrases in terms of a cut in the world, although 
we will follow Wigner's phrasing of this problem, below, instead.  That is, Bell
points out that the usual way of using the axioms of Quantum Mechanics involves 
introducing a cut in Nature, on one side of the boundary, we apply the first 
three axioms, and on the other side of the boundary, we apply the second three 
axioms.  Bell claims that this is not logically precise until the position of 
the cut in Nature is rigourously specified by its own axiom, specified once 
for all and in advance \dots we prefer Wigner's formulation of the illogicality 
involved, which Wigner calls a dualism instead of a cut.  The second point is 
that the measurement processes should be either defined, or primitive, but 
not both\dots )

Notice that the orthodox interpretation of the wave function's *values* at a 
particular point (as being related to the probability of finding the particle 
at that point in space or whatever) is *not* part of the axioms.  Indeed, 
when Born first introduced this interpretation, it was not immediately accepted:
the founders of Quantum Mechanics had been working successfully with the axioms 
alone and without any interpretation of the wave function except what we have 
given here, that it represents the physical state of the system.
Taking Bell's point seriously, one must decide whether the Born interpretation 
is an interpretation or a theorem.  It is known that it follows from the measurement 
axioms, so if we succeed in showing that the measurement axioms follow as 
approximations from the first three axioms, we will have clarified the status 
of the Born interpretation as well, showing that if it is not assumed, nevertheless 
some approximation to it follows as a consequence of the usual procedures of 
Classical Statistical Mechanics.

 
Wigner's discussion has the merit of isolating a purely logical question,
which has physical significance, and no excess philosophical baggage.
We will, in recalling his discussion, set up notation  
which will be used throughout this paper. 
 
Consider the physical situation underlying a measurement process which obeys these axioms. 
There is a `microscopic' system, or `incoming particle'---we use these terms purely for  
mnemonic convenience, for us they have no conceptual significance.  In practice, the system  
being measured is usually a photon or an electron approaching the apparatus from the left, say, 
and if it were a closed system not interacting with the apparatus, it would be thought of as  
a particle described by a wave function of its own.  If, then, it were a closed system,  
by axioms 1-3 it would be described by a wave function $\psi_o$ in a Hilbert Space $\Cal H_o$ 
with an intrinsic, time-independent Hamiltonian operator $H_o$.  Here and from now on, the  
subscript ``nought'' refers to the microscopic system being measured, not to an initial time or anything like that. 
From now on for definiteness we refer to the apparatus  as an amplifying  
apparatus so as not to prejudge the question of measurement.  
Now the amplifying apparatus would, if it were a closed system,  
also have its own mathematical objects, $\psi_m$, $\Cal H_m$, and $H_m$.  Of course some states of  
the apparatus are suitable for detecting the particle, 
 and others are not.  For example, a Geiger  
counter may have just been discharged, and unable to detect.  Or unplugged, or broken \dots. 
We suppose that $\psi_m$ is a state where it is ready to detect.   
 
We no longer believe that there is a cut which divides the quantum world from the classical world, 
so we now believe that the amplifying apparatus itself is subject to axioms 1-3, as  
indicated, even if it is `macroscopic' in the common-sense meaning of the word.   
Furthermore, the axioms seemingly apply to the joint system as well.  
As follows: as usual in Quantum  Mechanics, the Hilbert space describing the joint system
 is just  
$$\Cal H^c = \Cal H_o \otimes \Cal H_m$$ 
where the superscript $c$ stands for ``combined.'' 
If the microscopic system is in the state $\psi_o$ and the apparatus is in the state $\psi_m$, 
then the combined system is in the state given by the wavefunction $\psi_o \otimes \psi_m$.   
If there were no interaction, the dynamics of the joint system would be given by the joint  
Hamiltonian given by  
$$H^c = H_o \otimes I + I \otimes H_m$$ 
where $I$ means the (appropriate) identity operator. 
Since there is an interaction, then, tautologically, we have that the Hamiltonian of the joint  
system can be written  
$$H^c = H_o \otimes I + I \otimes H_m + H^{int}$$ 
where $H^{int}$ is a linear operator on $\Cal H^c$ which is thought of as the interaction term. 
 
Wigner pointed out what he called a dualism in Quantum Mechanics: the same physical set-up which  
we are discussing can be analysed in two different ways, and although there is no logical  
contradiction or disagreement between these two ways, that is only because there is no way to  
compare them, either.  Since the joint system is, at time $t=0$, in the state  
$\psi_o \otimes \psi_m$, the joint system is, at any time $t$, necessarily in the state  
described by  
$$\psi_{t} = e^{2\pi (\frac thH_o\cdot\otimes I +\frac thI\otimes H_m +\frac thH^{int})}\cdot (\psi_o\otimes \psi_m).$$ 
 
On the other hand, the amplifying apparatus is fitted with a dial, labelled with the eigenvalues  
0,1, say, for spin up and spin down (or vice-versa) and has a definite probability for  
being in one or the other.   
 
The only reason these two analyses have not led to contradictory results is that there is no way  
to interpret the one in terms of the other, and hence no way to compare them.  There is experimental support for both.  Interpreting the probabilities as frequencies, as usual in classical Statistical Mechanics, the measurement axioms are extremely well supported provided only that one has a practical sense of when something is a measurement and when not.  Recent developments in technology are eroding our sense of this divide: it used to be clear that measurement apparatuses were macroscopic and quantum systems were microscopic, one never observed peculiarly quantum effects such as superposition and entanglement except for microscopic systems.  But with the advance of mesoscopic engineering and quantum teleportation, this dividing line, really it was a demilitarised zone, is finally being populated and explored, so the usual practical sense is less of a useful guide in this regard.  The axioms involving Schroedinger's equation are very well  
verified, experimentally, so much so that it would be an act of desperation to introduce  
changes in the equation (such as have been proposed) such as non-unitarity, non-linearity,  
or stochasticity merely in order to solve the problem of Quantum Measurement and not based on experimental observation of new forces or effects.  
 
One way of looking at the problem, a less helpful one than the one we will adopt, has been  
to say that the first three axioms describe a linear deterministic dynamics which applies to all  
systems as long as they are not observed or measured.  Or, as long as the systems undergo processes which are not measurement processes.  They govern all physical processes except measurement processes.  The second triad of axioms govern measurement processes and are non-linear, discontinuous (approximately) and stochastic.  This way of looking at the problem sneaks in some unwarranted assumptions.  This will become clear if we take Einstein's point of view (minus whatever notion he may have had about hidden variables, which are now, and rightly, regarded as against our physical intuition, and so are rejected by the program of this  
paper).  In classical Statistical Mechanics, one had deterministic dynamical axioms, and from  
these, by means of purely mathematical methods, one derived probabilistic approximations. 
Our program, then, is to derive the three stochastic measurement axioms from Schroedinger's  
equation without making any new postulates, simply using statistical mathematical methods of  
analysis of a quantum system with a large number of component parts and degrees of freedom. 
 
From this point of view, the dualism Wigner points to is a gap, the lack of a definition of measurement process in terms of the first three axioms, and even more important, the lack of a definition of the phrase ``the result of the measurement process is $\lambda_i$.''We would like to say that the measurement apparatus, call it $M_\infty$, possesses a pointer variable $f_\infty$. Probably $M_\infty$ will be modelled by a mathematical space of some sort and, if so, then probably $f_\infty$ will be some sort of function (or maybe an operator on a space of functions) defined on $M_\infty$.  (The mathematical nature of $M_\infty$ and $f_\infty$ must remain further unspecified in order to avoid introducing unwarranted assumptions.)  We would probably like to have that the statement ``the result of the measurement process is $\lambda_i$'' is modelled by the behaviour of $f_\infty$, perhaps by its taking the value $\lambda_i$ or something like that, but these desiderata have to remain rather vague, because their further specification involves making various physical and  
mathematical committments, and we need to keep these committments separate from each  
other and give each successive commitment explicit scrutiny.  A later section of the paper will  
be devoted to analysing, at this level of abstraction, the classical idea of the thermodynamic  
limit in the works of Darwin--Fowler, Khintchine, and Ford--Kac--Mazur.

Wigner himself makes the experimentally unwarranted but philosophically tempting assumption, quite explicitly, 
which he calls, `psycho-physical parallelism' and von Neumann, after him, does the same.  For  
only this reason, Wigner goes further in specifying the problem, and makes it insoluble.  He  
assumes that our perception of the macroscopic measurement apparatus's giving the result  
$\lambda_i$ must correspond to a wave function of the apparatus or a more or less well-defined  
set of wave functions.  
Furthermore, Einstein's program  
suggests the opposite, since the thermodynamic limit of a sequence of dynamical systems can  
be, and usually is, a dynamical system with a totally different state space, a state space  
which is not in any obvious sense a limit of the other state spaces. 
 
Wigner, instead, goes on to further concretely specify the problem as one of entanglement. 
Suppose the incoming particle is described by a two-dimensional Hilbert space spanned  
by orthonormal wave functions $\psi_\varepsilon$.
 
Assume that the particle has a spin up state described by the wave function $\psi_1$  
and a spin down state described by $\psi_o$.  Assume that the measurement apparatus,  
when plugged in, calibrated, charged up and ready to detect the particle, is in state  
$\varphi^m_o$.  Wigner postulates that if the particle is in state spin up, then the  
apparatus moves, after a time period, its pointer to point to `u' and its wave  
function is then $\varphi_u^m$ whereas if the particle was in state spin down, then  
the apparatus evolves to $\varphi^m_d$, say.  Hence the joint system if in the state  
$\psi_o\otimes \varphi^m_o$ evolves after unit elapsed time to $\psi_o\otimes  
\varphi^m_d$ but if in the state $\psi_1\otimes \varphi^m_o$, to $\psi_1\otimes  
\varphi^m_u$.  Wigner points out that the linearity of Schroedinger's equation 
forces that a superposition of states on the part of the particle leads to entanglement. 
That is, if $v=c_1\psi_o + c_2\psi_1$, then $v\otimes \varphi^m_o$ evolves to  
a state which cannot be written as $\psi\otimes\varphi^m$ for any choice of states of  
the particle and the apparatus.  Such tensor products are called decomposable, and for  
them, and only them, do the particle and the apparatus have separate identities.  What  
we actually get is $c_1\psi_o\otimes\varphi_d + c_2 \psi_1\otimes\varphi_u$ and  
this is called an entangled state.  The axioms of quantum mechanics actually forbid us  
to interpret this as if it were a classical mixture, as if it meant the joint system  
had a probability $\vert c_1\vert^2$ of being in the decomposable state $\psi_o\otimes  
\varphi^m_d$ and a probability of $\vert c_2\vert^2$  
of being in the decomposable state $\psi_1\otimes  
\varphi^m_u$.  To go somehow from this analysis *to* the forbidden classical  
mixture interpretaion has been called by J.S.Bell, ``the Philosopher's Stone of Quantum  
Measurement.''  But to pose the problem in precisely these terms is to commit the fallacy  
of misplaced concreteness. 
 
It is a misplaced concreteness to assume that pointer positions are well defined 
sets of wave functions of the apparatus.  We will see that, granting that the 
measurement axioms are only approximately valid for constructible amplifying 
apparatuses obeying the laws of reality, this fallacy is much the same as assuming that 
the only approximations which have physical validity are those describable in 
terms of the norm topology.  But the approximations of classical statistical 
mechanics, which Einstein had in mind, do not fit into this misplacedly concrete 
paradigm.

In summary, we do not adopt the exact viewpoint of some on the problem.  We do not say that  
we have to define when the linear dynamics is valid and where the cut is which determines  
when the stochastic non-linear dynamics becomes valid.  Instead, we wish to derive the second  
three axioms (or approximations thereto) from the first three axioms.  The approximations,  
as usual, will be good approximations within a certain domain of parameters, and less good  
or even downright bad, elsewhere.  That is the point.  We wish to give a rigourous definition 
in physical terms and in terms of the behaviour of its Hamiltonian, of which physical processes 
are measurements and which are not.  The definition will be, that a physical process is  
a measurement when the sort of approximations we will derive are useful descriptions.  This  
is totally non-subjective and non-circular.  In principle, this will lead to experimental  
consequences as mesoscopic engineering develops, which can discriminate between this  
solution and others which have been proposed.  In particular, we accept that entanglement  
persists.  Our model provides, in principle, a theoretical basis for determining which  
observables will be able to detect macroscopic superpositions and entanglements, and which  
will not be able to detect them.

Wigner's paper's formulation of the problem has already been outlined.  We outline  
his solution and show why it does not quite fit into our framework.   
 
Wigner has proposed that the definition  
of measurement process is that a consciousness is involved, and that when a consciousness is  
involved, in reading off the position of the pointer on the dial, then a new, still  
undiscovered non-linear dynamical equation governs the process.  He has not generally been  
followed in this.  His solution does not fit into our framework because he does not derive  
this non-linear equation from the linear one.  But his acceptance of the philosophical dualism  
that separates consciousness from unconscious matter, a dualism that goes back to Descartes  
and Malebranche and Leibniz in modern times, is extemely interesting, and will have a role  
to play in philosophy. 
 
Wigner assumes without proof that the measurement apparatus must have pointer positions that  
are describable by wave functions.  The axioms do not assert this.  Experimental practice would  
be unchanged if this assumption were abandoned or contradicted.  There is a vast difference  
between `interpretation' and `operationalisation' and we will change the interpretation of  
pointer position from Wigner's without changing its operationalisation in terms of laboratory  
procedures.  The axioms do not force Wigner's interpretation (which is the usual one) on  
us since the axioms do not interpret themselves.  The phrase, ``the result of  
a measurement process'' is, just as is the concept of `measurement,' an undefined one, a  
primitive one.  Operationalisation is the way we turn physical concepts or the mathematical  
models of them into concrete laboratory procedures.  If one adopts a particular operationalisation 
this may well impose various constraints on the interpretations that can be consistent with  
it, but it by no means determines them uniquely.  (And vice-versa, of course.)  We will end up 
by abandoning the philosphical assumption of psycho-physical parallelism in favour of a much  
looser but still physically precise correlation.  This would involve modifying the Cartesian  
dualism {\it cum} parallelism to a dualism between mind and matter which is not precisely  
parallel, but admits of various rigourous correlations which are not one-to-one mappings. 
This is just as consistent with the experimental evidence as is Wigner's philosophical  
assumption.  The assumption of psycho-physical parallelism, which is a philosophical  
assumption, is a fallacy of misplaced concreteness.

Of course if we accept Einstein's program as valid, we have to accept that the measurement  
axioms are approximations to physical reality analogously to the status, in classical  
physics, of the approximations of Statistical Mechanics such as temperature and phase  
transition, which are not exactly valid for finite systems but only are well-defined in  
the thermodynamic limit.  Probabilistic formulas are approximations to deterministic  
reality. 
 
It should be pointed out that once we accept that the measurement axioms may be satisfied  
approximately instead of exactly, there are indeed many kinds of approximation: there is  
no experimental evidence to force us to choose the strong topology.  (In this paper, we  
are going to adopt the same sort of approximation as used in the classical statistical  
mechanical papers of Darwin--Fowler and others.  Coleman--Hepp adopt the same sort of  
approximation as used in Quantum Statistical Mechanics and the theory of local algebras 
and inductive limits of $C^*$-algebras.)   Some of Bell's criticisms  
are misguided on this issue, he reasons as if only the strong topology is acceptable. 
But there are reasons why the strong topology on the space of observables cannot possibly  
be of physical significance for measurement, based on the Araki-Wigner-Yanase theorem 
(that observables with eigenvalues closer and closer together require larger and larger,  
without bound, measurement apparatuses to measure them with a fixed degree of accuracy). 
We will discuss this issue in the next subsection.

Bohr proposed that measurement apparatuses were classical.  We  
explicitly implement this old idea in what is a novel way: we find a thermodynamic limit  
of quantum systems which is a classical system.  This is in sharp contrast to Coleman-Hepp  
and every other statistical mechanical approach to the problem of Quantum Measurement, which  
studies a thermodynamic limit system which is still quantum.
Bohr on numerous occasions asserted (without proof) that the measurement apparatus and  
pointer positions implicit in the axioms had to be a classical apparatus, or at least  
classically describable.  In conversation with Rosenfeld, he once explicated this further: 
the apparatus has to be such, that its pointer positions are such, that it is a valid  
approximation to neglect Planck's constant, that is the precise meaning he was pushed to  
giving to his more famous phrase, ``classical in nature.''  We implement this by introducing  
a renormalisation in our limiting process that sends Planck's constant to zero.   
As always (prior to the $C^*$-algebra approach), the thermodynamic limit itself is unphysical,  
a mere mathematical convenience for easily obtaining good approximation to a physical system  
which has a large but finite number of degrees of freedom. For an actual amplifying apparatus 
with a large but finite number of components, and a small but non-zero value of Planck's  
constant, the measurement axioms calculated by considering instead the limit as the number of  
components grows without bound and as Planck's constant decreases to zero without an upper  
bound to its reciprocal, are a good approximation for the purposes of pointer positions  
(although not, say, for irrelevant aspects of the amplifying apparatus such as the radium  
paint foolishly used to decorate the dial\dots).  But there are other considerations which  
suggest that the measurement axioms are exactly valid only about an unphysical thermodynamic  
limit of some sort, besides the correspondence principles and Bohrian tradition.  These  
will be discussed separately in the next subsection.

 The goal of this paper is 
to show that Einstein's program can be accomplished in a way hitherto assumed, 
without sufficient proof, to be impossible.  We show, by using a more  
foundational approach to classical statistical mechanics, that hidden variables are 
unnecessary.  
Referring to the list of desiderata and avoiderata in ``Against Measurement'', this paper avoids 
 all use of the banned terms environment, reversible, irreversible, information, measurement. 
 And, I might add, dissipative, decoherence, random, mixture, density matrix.  
(For us, all systems are closed systems, all states are pure states, and all superpositions  
are coherent, always.) The banned terms 
 system, apparatus, microscopic, are merely used as mnemonic labels and are not relied on to  
derive substantive conclusions. The banned terms 
 macroscopic, and observable  are precisely defined in terms of acceptable notions.   
By observable, we will mean an abelian 
observable as in classical Hamiltonian mechanics, and we give this a precise meaning  
in terms of the axioms of quantum 
mechanics.   
The usual primitive notion of quantum mechanical observable is not used.  Nor do we study  
linear operators and their eigenvalues.  The Hamiltonian is not regarded as an observable  
at all, but simply, as in classical mechanics, as the infinitesimal 
generator of the dynamics.  We derive the usual probabilistic axioms about the results of  
measurements from the axioms which do not involve the concept of measurement or observable.   
In particular, although the whole approach of this note is statistical, we do not make any  
statistical hypotheses.  We make only deterministic hypotheses as is usual in any Hamiltonian  
dynamical system, but use statistical methods to study those systems.   
 This will become clearer by example.  
 
Schwinger has written an important paper on the subject of quantum Brownian motion. 
He does not make any connection to the topic of Quantum Measurement.  Nevertheless,  
there are important results claimed (but not proved explicitly) in the paper which  
turn out to be relevant.  He derives formulae for the interaction of a microscopic  
particle with a Hamiltonian heat bath at a negative temperature, which produces  
a negative temperature Brownian motion in the particle.  He remarks that this  
motion amplifies the quantum motion of the particle to the classical level, where  
quantum uncertainties and non-commutativities are negligible.  Hence, every key  
idea of this paper (writing down an explicit Hamiltonian for an amplifying apparatus may  
not have been done before, but it is routine) has been anticipated except one, the  
definition of macroscopic observable. 
 
The self-restraint of the adopted program of this paper rules out the postulating of 
open systems, since they are not allowed by the axioms.  By definition, in the foundations  
of the classical epistemological approach to physics, system means closed system.  The  
axioms are for closed systems only.  Nevertheless, something should be said in this  
section about the open systems approach to the problem of Quantum Measurement. 
Zurek \it et al.\ \rm and Zeh \it et al.\ \rm among many others are prominent proponents  
of the open systems approach, which makes the peculiar (from the point of view of Wigner  
and the axioms) properties of Quantum Measurement depend upon the interaction of the  
joint system of microscopic system being measured \it cum \rm measurement apparatus  
with the environment in a way reminiscent of the phenomenological approach to Hamiltonian  
heat baths, Brownian motion, and other thermodynamic limit phenomena.  Up to now, they  
have no experimental support for their proposals, and also have phenomenological  
parameters which they adjust in order to achieve this.  They merely postulate a master  
equation of some sort, but do not derive it from an underlying deterministic dynamics. 
This is entirely analogous to the difference between the Langevin approach to Brownian  
motion and the Wiener approach.  Wiener's approach allows us to derive the stochastic  
differential equation from Newton's equations, as shown by Ford--Kac--Mazur in an  
immensely influential paper.  Wiener's approach is adopted here, in imitation of the  
beginning parts of that paper.  Bell would criticise the open systems approach because  
of its use of undefined notions such as decoherence, and this is a foundationally important  
issue to clear up: it is important to know whether or not the problem can be solved  
without introducing new primitive notions.  But whether Nature uses one solution or the  
other cannot be specified a priori by foundational desiderata.  Although from our viewpoint it  
seems counter-intuitive to make the reduction of the wave packet, or the decoherence of  
the entanglement of the pointer pointing to zero with the pointer pointing to 1,  
a macroscopic superposition of states, depend on interaction with a thermally stable  
environment coupled by only relatively weak forces to the apparatus,  this is a  
question for experiment.  Recent success in quantum teleportation shows that  
entanglement persists over dozens of miles with no specially quantum efforts to screen the  
system from interaction with the environment.   
 
This means that it is up to future experiments to determine whether the peculiar behaviour of  
measurement apparatuses is due to their interaction with the environment, or due to their  
interaction with the system being measured.   
 
\centerline { The Paradox of Degeneracy} 
 
In this sub-section we adduce new considerations, a kind of thought experiment in taking  
the six axioms completely literally, which suggest two things: the norm topology on  
operators is not physically significant for observables, and, the notion of observable  
might be the result of a limiting procedure of some sort, only approximately true in  
any finite-sized laboratory.  We connect this with results of Araki-Wigner-Yanase. 
The similarities with the status of phase transitions in classical Statistical Mechanics  
is striking, and suggests that the mathematical behaviour of an observable should be  
derived by a mathematical limiting procedure (such as the thermodynamic limit) from  
physically realisable phenomena. 
 
The principle of continuity has a long history in Physics, and underlies the notion of  
a real variable.  It states that small variations in the physical set-up of the lab  
should produce only small changes in the experimental results.  This has implications  
for the mathematical model used to model the physical set-up and the experimental results. 
It might be thought that this principle has been discarded by Quantum Mechanics, but this  
is not so.  Dirac, for example, appeals to it in justifying the axiom of the reduction  
of the wave packet, in his analysis of repeated measurements.  The impression that  
the principle is refuted or discarded is based on a misunderstanding of what are the  
experimental results in Quantum Mechanics, in the real world.  No particular quantum  
jump or discontinuity is ever predicted by the calculations of Quantum Mechanics, only  
probabilities or, equivalently, expectation values.  These probabilities do indeed vary  
continuously with continuous variations in the physical parameters of the lab set-up. 
More contortedly, the observation that the measurement process resulted in $\lambda_i$ 
``this time'' is not, strictly speaking, an experimental result.  Experimental  
results are, by definition, replicable, and this is not replicable (except in the  
special case that the probability was unity).  Observations that are replicable include 
``it can happen that the result is $\lambda_i$,'' or, ``the probability that the  
result is $\lambda_i$ is in between .1 and .15,'' and things like that.  The result of  
a single physical process, occurring once, is not an experimental result unless it  
is replicable.  Nature and Heisenberg have taught us that only the probabilities  
are replicable, therefore, only the probabilities are experimental results.  If  
Heisenberg taught us anything, it is that a physical theory should not be criticised  
except for disagreement with experimental results, and it follows from this that  
a physical theory does not have to explain more than the probabilities.  Therefore the  
principle of continuity only applies to continuous variation of probabilities and  
expectation values with continuous variation in physical set-up.  There is no reason,  
yet, to suppose that Quantum Mechanics violates the principle of continuity.  But  
in the following thought experiment, it will be seen that the mathematical model of  
the measurement axioms does indeed violate the principle of continuity in a hitherto  
unsuspected way.  We are going to construct a family of observables which vary  
continuously in the norm topology, but yield an experimental set-up where a certain  
expectation value varies discontinuously. 
 
 

\def\[{\lbrack}\def\]{\rbrack} 

\loadeusm \font\scr=eusm10

\def\foo#1#2{\baselineskip=10pt\plainfootnote{$^{#1}$}{#2}\baselineskip=24pt\thinspace}    \def\ep{\epsilon} \def\sh{\text{\scr H}}


 \

 Briefly, if $Q$ is an observable with degenerate  eigenvalues and if  $Q_\ep$ is a family ($\ep>0$) of perturbations  of $Q$ which remove the degeneracy, then we consider the  reduction of the wave packet produced on the one hand by  measurement of $Q$ and on the other by measurement of $Q_\ep$ as    $\ep$ approaches zero.  We obtain a discontinuous variation in  the expectation value (of a different observable) as a  function of $\ep$. This violates the principle  of  continuity.\foo1{ This is the \it only \rm violation of this principle in Quantum Mechanics.  The so called `quantum jumps' are not discontinuous, except in slang.  For only functions can be continuous or discontinuous.  But it is precisely in the conventional interpretation of Quantum Mechanics that the result of an experiment is not a function (of anything).  The probability with which a given result will be observed is a function of experimental conditions---but this depends continuously on the physical parameters, except in the case to be discussed in this paper.}

  However if one interprets the reduction of the wave  packet as an approximation in a specific way, it is  predicted that this variation should be smoothed out.   

 Let $\sh_1=\bold C\cdot\vert\uparrow\rangle\oplus\bold
C\cdot\vert\downarrow\rangle$ be a two dimensional Hilbert space      which is
the state space of particle one.  Let $\sh_2$    be  isomorphic to  $\sh_1$  
and be the state space of a  distinguishable particle, particle two.  Let 
$\psi=|\uparrow\rangle\otimes|\uparrow\rangle+|\uparrow\rangle\otimes|\downarrow\rangle$. 
Let $Q$ be the (degenerate) observable `` `spin' of the  first particle," taking
eigenvalue 1 if the first particle  has spin up and 0 if spin down.
 Now the eigenvalues of $Q$ each have a two-dimensional degeneracy.  Let $Q_\ep|\uparrow\rangle\otimes|\uparrow\rangle=(1+\ep)|\uparrow\rangle\otimes|\uparrow\rangle$     and $Q_\ep|\uparrow\rangle\otimes|\downarrow\rangle=(1-\ep)|\uparrow\rangle\otimes|\downarrow\rangle$  (and  be otherwise unperturbed).  This perturbation removes the  (relevant) degeneracy.  We suppose the physical system is  in state  $\psi$.  The results of experiments $Q_\ep$ pass  continuously to those of $Q$  as $\ep\rightarrow 0$.  

 But the reduction of the wave packet does not.  If $Q$   is measured, $\psi$  is unchanged.  If $Q_\ep$ is measured, then  there are equal chances that $\psi$  jumps to $|\uparrow\rangle\otimes|\uparrow\rangle$  or that it  jumps to $|\uparrow\rangle\otimes|\downarrow\rangle $  but it never lands anywhere else in the  span of those two vectors.  Let $R$ be an observable which  is zero on the zero-eigenspace of  $Q$, and which has as  eigenvectors  $|\uparrow\rangle\otimes|\uparrow\rangle+ |\uparrow\rangle\otimes|\downarrow\rangle$               and $|\uparrow\rangle\otimes|\uparrow\rangle- |\uparrow\rangle\otimes|\downarrow\rangle$, with eigenvalues one and zero, respectively.

 If $Q$  has been measured, then the expectation value of  $R$  is unity.  But if $Q_\ep$   has been measured for any  \ $\ep>0$, the  expectation value of $R$  is half of that.  \  This is a violation of 

 \ \ 

\vskip 1.6 in

\qquad\qquad
\qquad\qquad\quad\includegraphics{Fig1.eps}

 \noindent the principle of continuity which, indeed, is not an axiom of Quantum Mechanics, but  seems physically warranted in this example.

 It  is  well-known\foo1{Araki, H., and Yanase, M.,  Physical Review, vol. 120, p.\ 622 (1960).}  that the  performance of a measurement that discriminates with  increasing precision between eigenvalues that approach  more and more closely together requires increasing  resources (laboratory size, etc.)  It would seem  therefore, that the passage from $Q_\ep$  to $Q$  is not physically  continuous even though it is continuous in operator norm.

 One might hypothesize that the degree of validity of the approximation of the  reduction of the wave packet rule to the behaviour of the actual, finite size, physical apparatus, depends on three things which affect the convergence:  the  exact initial condition  $\psi$, the magnitude of the apparatus, call it $n$, and the distance between neighbouring eigenvalues.  Even if we neglect the  first two of these, it is obvious that the third is  variable as  $\ep\to0$   and so the variation of the expectation  value will be smoothed out, as is easy to see.  

\ \ \ 

\vskip 1.6 in
\qquad\qquad\includegraphics{Fig2.eps}

Benatti\foo2{Personal communication} has worked out a derivation of the smoothing effect from the general assumptions of 
the Ghirardi--Rimini--Weber approach to quantum measurement, which involves replacing 
Schroedinger's equation by a stochastic differential equation.

Finally, we remark that it seems to be a common mistake to think that because $\psi(x,t)$ is not continuous, that therefore the principle of continuity does not hold in Quantum Mechanics.  Firstly, for fixed $t$, $\psi$ is not a function, it is an element of $L^2$ and its value at a point is not defined, and so is certainly not physical.  Rather, $\int_{x_0}^{x_0+\epsilon}|\psi(x)|^2 dx$ is physical---but it is also a continuous function of $x_0$ and $\epsilon$.  Secondly, for variable $t$, although it is true that $\psi$ has a jump discontinuity when a measurement is made, it is also true that precisely in the usual interpretation of Quantum Mechanics, because it is non-positivistic, $\psi$ is not regarded as objective and physical: only the results of calculations using $\psi$ in order to predict results of experiments are.  And here, as remarked above, there are no discontinuities \it except \ \rm in the case of a degenerate observable.

     Even so, the fact that, at least in the $C^*$-algebra context, a measurement theory based on taking the thermodynamic limit of the interaction between a (fixed) microscopic system and a (variable) macroscopic measuring device has been constructed is encouraging. Classically it is of course well known that for any finite point at which we stop short of the thermodynamic limit, we have determinism.  Only at the unphysical idealized thermodynamic limit do we obtain probabilistic behaviour.  So it seems likely that the same thing is true for quantum systems.  This would suggest, as a dependence on $n$, the same smoothing effect we have predicted (as a function of $\epsilon$).

     It also suggests that the principle of continuity will hold at every finite point, and that the discontinuities usually taken to be part of the structure of theoretical physics are simply an unphysical artifact of the approximation procedure inherent in passing to the thermodynamic limit. 

\vfill \eject

\centerline {3. \bf The Concrete Model of the Ming Effect} 
 
From now on, we distinguish between measurement apparatus and amplifying apparatus.  The  
amplifying apparatus we study will be an explicitly given quantum system with $n$ degrees  
of freedom, $M_n$, modelled by a Hilbert Space $\Cal H_n$ and with a Hamiltonian $H_n$. 
It approximates more and more to a measurement apparatus as $n\rightarrow\infty$.  The  
measurement apparatus is a thermodynamic limit of $M_n$, denoted $M_\infty$, and is a  
classical dynamical system.  Its states are the equilibrium states of the thermodynamic  
limit, and are not described by wave functions, its state space is a symplectic  
manifold, not a Hilbert space, has no linear structure, superposition of states is a  
nonsensical undefined concept for it.  Classical mixtures of its states are possible,  
as always in classical Statistical Mechanics.  One can take the viewpoint that  
measurement apparatuses and processes are unphysical idealisations of the only processes  
that are physical, the amplifying processes.  This is a valid logical interpretation of  
the measurement axioms (even, after some contortions, the reduction of the wave packet) 
and it does not involve any change in the operationalisation of the concept of  
measurement.  In fact, it grounds in concrete calculations what used to be operationalised  
anyway without justification: the fact that an amplifying apparatus must be large before  
the measurement axioms are verified.  A one-atom device does not perform a measurement\dots  
or reduce the wave packet. 
 
Let the state space of an incident particle be $\bold C^2$.  This space has 
basis $\{\psi_0, \psi_1\}$.   For each $n$,  
$\Cal H_n$ is the Hilbert space of wave functions describing the state space of an  
$n$-oscillator system which is an amplifying device.   
We let $\Cal H_n=\oversetbrace {(n)} \to {\bold C^2\otimes\bold C^2\dots\bold C^2}$.   
 
 \vskip -.14in   In the presence of an incident particle in the state $\psi_1$, the amplifying apparatus will  
 evolve in time under the influence of $A_n$ (called ``Ming,''since it leads to a bright and clear phenomenon), a  cyclic 
nearest-neighbour interaction which is  
meant to model the idea of stimulated emission or a domino effect.  In the absence of a  
detectable particle, 
 the dynamics on the amplifying device will be trivial.   
This means that  
 
\vskip -.14in $\Cal H_n^{com}=(\bold C\cdot  \psi_1 \otimes \Cal H_n) \bigoplus (\bold C\cdot \psi_o \otimes \Cal H_n) 
 \text {\ \ and we put \ \ } H_n^{com}=I_2\otimes A_n + I_2\otimes I_{2^n}.$  
The intuition is that $\psi_1$ means the particle is in the state which the apparatus is designed to detect.   
 but $\psi_o$ means the particle is in a state which the apparatus is designed to ignore.

\vskip -.16in It will simplify things if we assume $n$ is prime.  (The general case can be reduced to  
this by perturbation.) 
Let $q={2^n-2\over n}$ (which is an integer by Fermat's little theorem). 
 
\vskip -.07in Let $i$ be any integer between 0 and $2^n-1$.  There are $n$ binary digits $d_i$ with  
$i=\sum_0^{n-1} 2^{i}d_i$ and uniquely so.  If $\{\vert 1\rangle,\vert 0\rangle\}$   
is a basis for $\Cal H_1$, then $\vert i\rangle_n= 
\otimes_0^{n-1} \vert d_i\rangle $ form a basis of $\Cal H_n$.  The intuition is that the  
$i^{th}$ oscillator is in an excited state $\vert 1 \rangle$ if $d_i=1$ and is in the  
ground state $\vert 0\rangle$ if $d_i=0$.  We also write  
$\vert i \rangle_n=\vert d_0 d_1 \dots d_{n-1} \rangle$. 
 
We wish to find Ming such that in one unit of time the  
$d_i$ are cycled as follows: 
 
\vskip -.46in $$e^{\frac{2\pi} h  A_n} \vert i \rangle_n = \vert d_{n-1} d_0 d_1 \dots d_{n-2} \rangle.$$

\vskip -.28in Choose a set of representatives $b_i$ such that every integer $k$ from 1 to $2^n-1$ can be written uniquely as 
$b_i2^m \mod (2^n-1)\bold Z$ for some $1\leq i \leq q $ and $0\leq m\leq n-1$, that is, $k=b_i2^m +j(2^n-1)$ for some $j$ 
but $i$ is unique.  (Since $2^n-1 $ and $2^m$ are relatively prime, no matter how $k$ and $m$ are fixed, there exist  
unique $b_i$ and $j$ satisfying this.)   
 
Then each $|b_i\rangle$ represents an orbit under the action of $e^{\frac {2\pi} h A_n}$.  Re-order the basis as follows: 
let $v_o=|b_1 \rangle$, $v_2=|b_12\rangle$, $v_2=|b_12^2\rangle$, \dots $v_{n-1}=|b_12^{n-1}\rangle$, 
$v_n=|b_2\rangle $,  
$v_{n+1}=|b_22\rangle$, \dots, $v_{2n-1}=|b_22^{n-1}\rangle$, $v_{2n}=|b_3\rangle$, etc., up to 
$v_{(q-1)n}=|b_q\rangle$, 
$v_{(q-1)n+1}=|b_q2\rangle \dots $,  
$v_{(q-1)n+n-1}=|b_q2^{n-1}\rangle$,  
but $(q-1)n+n-1=2^n-3$, so we have $2^n-2$ basis vectors accounted for.  Let $v_{2^n-2}=|0\rangle$ and 
$v_{2^n-1}=|2^n-1\rangle.$ 
 
Let $V_1$ be the space spanned by $\{v_o,\dots,v_{n-1}\}$, let $V_2$ be the space spanned by $\{v_n,\dots,v_{2n-1}\}$, 
etc., up to $V_q$.  Let $V_o$ be the space  
spanned by $\{v_{2^n-2}, v_{2^n-1}\}$.  
The Ming Hamiltonian operator $A_n$ on $\Cal H_n$ is a direct sum of its restrictions to the $V_i$.  Its restriction to $V_0$ 
is to be the zero operator.  Each $V_i$ is isomorphic to $V_1$ and we give the matrix of each restriction of $A_n$ with 
respect to the given bases. 
$${\text {\hskip -1in Solving \quad \ \ \quad\quad\quad   \quad \quad \quad }}  
{A_n}=\frac h{2\pi}\log \pmatrix  
0 & {0}& \dots&{0}& 1\\ 
1&{0} & \dots&{0}&{0} 
\\ 0&1&\ddots&{\vdots}&\vdots 
\\ \vdots&\ddots&{\ddots}&0&\vdots 
\\  0 &{\dots}&0 & 1&0\endpmatrix  , $$  
we obtain a cyclic skew-hermitian matrix, 
whose $i,j^{th}$ entry,  
${-ih\over n^2}\sum_{k=0}^{n-1} ke^{{2\pi i \over n}k(i-j)}$, 
 is approximately (if $n$ is large compared to $i-j$) 
$ih{(i-j)^{-1}\over2\pi} {\text {\ \ unless\ \ }} i=j {\text{\ \ in which case \ }} {ih\over2}$.   
 
     As usual in classical statistical mechanics, the observables are all abelian, and are given by measurable functions on the 
phase space, hence $f_n$ is an observable if  
$f_n:\Cal H_n^{com} \rightarrow \bold R$ is measurable and $f(c\psi)=f(\psi)$ for $c\in \bold C^{\times}$.   
In order to avoid confusion with the orthodox primitive concept of observable, modelled 
by a linear operator, we will not refer to $f_n$ as an observable and will not use the 
term `observable' in our system at all.  These measurable functions are dynamical 
variables, as usual in Hamiltonian dynamics.

The intuitive picture is that this device is getting more and more classical as $n$ goes  
to infinity.  So the energy levels get closer and closer, approaching a continuum, the  
oscillators get closer and closer which is why the interaction, at a constant speed,  
travels from a oscillator to its neighbour in less and less time.  If we adjusted  
by rescaling the dynamics to accomplish this, the entries of $H_n$ would diverge with $n$. 
 We rescale $h$ instead, so that it  
decreases as $\frac 1n$, so that we have finite length and fixed density and fixed total  
energy.  This is typical of rescaling procedures in classical statistical mechanics.  It is physically meaningful because the 
thermodynamic limit is never physically real, it is  
only one of the $\Cal H_n$ which is physically real: $n$ is not a physical variable, it is a parameter.  Passing to the limit is only a 
mathematical convenience to obtain simple  
approximations for the physical truth about $\Cal H_{1.1\times 10^{23}}$.   
Since $h$ is truly small, this yields valid approximations. 

\centerline{\bf 4. Classical Statistical Mechanics} 
 
\vskip -.12in
The presentation of classical Statistical Mechnics in undergraduate, and even  
most graduate, textbooks is logically incoherent and is inadequate for the purposes  
of answering Bell's criticisms of the statistical mechanical approach to the  
problem of Quantum Measurement.  A logically coherent and adequate foundation  
for the subject was outlined, with many special cases done explicitly and compared  
to experimental results, by Darwin and Fowler in the 20's.  This approach was  
also taken up by Kolmogoroff, Khintchine, and Wiener, in scattered papers on the  
ergodic theorem.  Khintchine attempted a somewhat  
controversial generalisation of Darwin and Fowler's many examples, but we will  
not appeal to the controversial part of Khintchine's explanation, but only the  
unusually perceptive and acute formulation he gives of the foundations of  
classical Statistical Mechanics as a program.  In its full generality, it remains  
today as one of the challenges to pure mathematics to establish its scope of  
validity as a theorem, we do not need this. 

The notion of thermodynamic limit employed by these mathematical physicists does 
not seem to fit into the framework of topology.  These methods were acutely, if 
disparagingly, characterised by R.\ Minlos, ``For a long time the thermodynamic limit 
was understood and used too formally: the mean values of some local variables and some 
relations between them used to be calculated in a finite ensemble and then, in the 
formulas obtained, the limit passage was carried out.''  But this is much the same 
as to say that a method of double duality was employed \dots
 
The methods of Darwin--Fowler 
and Ford--Kac--Mazur are well supported experimentally.  The results of the quantum  
measurement axioms are well supported experimentally.  It adds to the likelihood of  
the correctness of our analysis that the statistical methods we will use in the next section
 are brain-damaged  analogues of well-supported methods and lead to well-known results.

It is also the case that these methods have been largely supplanted by $C^*$-algebra  
methods and the definition of thermodynamic limit as infinite system (not as a limiting  
process) of Ruelle and others.  But these latter methods are less flexible, so far.  
So far, they have only been used to study infinite volume limits.  But these are not  
the limits appropriate to Brownian motion.  Besides this, if we take Bohr seriously, 
we want an `infinitely rigid' limit, or `increasingly classical in nature' sort of  
limit, and this seems to involve smaller and smaller quantum units of action becoming  
located closer and closer to each other, nothing to do with infinite volume. 
Doubtless the classical limit of quantum systems we will construct in the next section  
could be fit into some kind of algebraic framework, but because of the paradox of  
degeneracy, it seems that the strong operator topology is un-physical,  
so it seems pointless to do so.   

The Gibbs program is to derive the probabilities from a deterministic dynamics 
via a Hamiltonian heat bath.  When applied to the question of Brownian motion, it 
takes the following form: for every integer $n$ we consider a Hamiltonian system 
of one Brownian mote and $n$ surrounding bath particles.  As $n$ increases, we 
may wish to let the mote remain at a fixed mass but let the other particles 
decrease in mass proportionately, keeping the total mass fixed.  Let each such
system be labelled $M_n$, it is a symplectic manifold as a phase space and 
possesses a deterministic flow on it given by the dynamics: given an initial 
condition $v_o\in M_n$, the system will be in state $v_o(u)$ after a period of 
time equal to $u$.  

They study a particular dynamical variable $f_n$ on each system: the momentum 
of the Brownian mote.  Or, rather, its auto-correlation function,
$$g_n(t)=\lim_{T\rightarrow\infty}\frac 1T \int_0^\infty f_n(v_o(u))f_n(v_o(u+t))du.$$
Or, rather, what would be equivalent if the system were ergodic, a phase 
average of $f_n(v_o(u))f_n(v_o(u+t))$, instead of a time-average.
That is, following Gibbs's original vision, they impose a Maxwell--Boltzmann 
probability distribution on the space $M_n$, supposing it to be a heat bath 
in equilibrium at a positive temperature.  
(This has been superseded in following papers.)  Now, a deterministic flow taken together 
with a probability distribution on the space of initial conditions yields a 
stochastic process, call it $P_n$.  This yields, in turn, a stochastic 
process on the momentum of the Brownian mote alone.  Hence for each integer $n$, they 
obtain a Gaussian stochastic process with auto-correlation function $g_n(t)$.
They pass to the limit as $n\rightarrow \infty$, which requires a cut-off and 
re-scaling procedure, obtaining a function $g_\infty(t)$ which they interpret 
as the auto-correlation function of a Gaussian stochastic process, even though 
they have as yet no phase space or probability space for the process to live 
on, much less the process itself.  They show that this limit function has no 
memory effects, even though the finite systems $M_n$ each satisfy Poincar\'e 
recurrence.
In fact, it corresponds to the Wiener process.
They have thus derived the Wiener process from placing a mote in a Hamiltonian 
heat bath at a positive temperature.

Let the stochastic process on each $M_n$ as above be $P_n$.  Let the Wiener 
process which arose in the limit be $P_\infty$.  If we regard their procedure 
as being somehow that the sequence of $P_n$ has, as a limit, $P_\infty$,  
then we have a notion of limit that does not seem to fit into the framework 
of topology very well.  It is very flexible.

Their paper has been very influential and further work has shown that the 
thermodynamic limit is largely independent of the particular probability distribution 
imposed on the bath, and robust with respect to the particular dynamics 
introduced on the $M_n$.  This is only to be expected, since 
thermodynamic limits are robust to the 
underlying dynamics, because of the central limit theorem.  In the next section, 
we will show that a stochastic process arises even without  a probability 
distribution being imposed on the bath.

 
In summary, then.
The procedure of Ford--Kac--Mazur was a kind of double duality.  Given a sequence of  
finite classical systems, $M_n^{com}$, they did not attempt  
to find the state space of the limit object from the state spaces of the sequence. 
Instead, they, before passing to the limit, passed to the  
consideration of dual objects,   (dynamical variables: of course the auto-correlation 
functions are a species of dynamical variables). 
Passing to the limit   
they obtained a candidate for a dynamical variable, 
 which they regarded as the dual object for an unknown 
dynamical system, to be sought.  We will carry  
this out in our setting (but using time averages for convenience). 
 
\centerline{5.\bf The Statistical Analysis of Amplification} 
 
Should we study quantum, non-commutative observables, i.e., linear operators, on each  
amplifying apparatus and then pass to the limit? For the reasons discussed in the paradox  
of degeneracy, the strong operator topology on observables, which in our model is the  
only relevant topology, cannot be foundational or of direct physical significance.  It  
probably arises as the result of a limiting procedure, so that the discontinuity in the  
paradox can be seen as a non-physical artifact of invalidly interchanging limits in a  
double limit.  Therefore we should study something else on each $\Cal H_n$ and derive  
the observable as a limit as $n\rightarrow\infty$.  Amazingly, the ordinary abelian  
dynamical variables suffice for this purpose, as if the space of rays in $\Cal H_n$  
were to be treated as a phase space exactly like in classical Hamiltonian Mechanics. 
Because only one measurement can be made at a time, that measurement commutes with  
itself, so abelian-ness is not a restriction.  The non-commutativity of the quantum  
observables will arise out of the fact that the different measurement apparatuses  
required to measure them force each sequence of abelian dynamical variables to live on  
totally disjoint spaces, so that their commuting with each other is a meaningless  
question.  If one and the same amplifying apparatus can measure two quantum observables,  
then they must commute.  But this was already pointed out by Bohr and Heisenberg.

\vskip -.08in
\centerline{ Macroscopy and pointer variables} 
 
\vskip -.12in
In order to do thermodynamics we consider sequences of observables  
which are abelian dynamical variables, $\{f_n\}_{n=1}^\infty$.  Each $f_n$ should have the 
same ``physical meaning'' relative to $\Cal H_n$.   
     It is unclear how to formalise this in complete generality.  If each $f_n$ is ``energy'' we are doing the same thing as 
Khintchine, Darwin--Fowler.  Ditto if each $f_n$ is a phase average of a measure of the fluctuation of energy of a single 
component.  Ford--Kac--Mazur consider the example where each $f_n$ is the momentum  
auto-correlation of one component.  (It is an important open problem, which is very difficult,  
to find the largest range of validity of this procedure, which they do in concrete examples,  
and give an abstract definition which will cover such a range.) 
 But we have to consider a sequence $\{f_n\}$ which captures the notion of ``visible to the naked eye'' or, ``macroscopic.''

At any rate, we formally define such a sequence of $f_n$ to be macroscopic if whenever the  
sequence of norm one vectors $v_i\in\bold C^2$, $i>0$, satisfies  
$$\lim_{n\rightarrow\infty} f_{n+n_o}(v_o\otimes v_{n_o+1}\otimes v_{n_o+2}\otimes\dots\otimes v_{n_o+n}) 
\text {\quad\quad\quad exists for some}$$ 
 $n_o$ and some $v_o\in\Cal H^{com}_{n_o}$, then it exists and is independent of the choice of 
$n_o$ and $v_o$.   
 
Many macroscopic observables, such as temperature, are not relevant.  Only macroscopic variables relevantly coupled to the 
incoming particle are pointer variables. 
 
     The concept of a pointer variable is one that had resisted precise definition.  The intuitive picture has always been that of a 
measurement apparatus which, after amplifying its response to the incoming microscopic system,  
makes a macroscopic needle point to a number on a dial in a naked-eye fashion.  If the  
measurement apparatus is classical this is easy to define, but then it is not so easy to link  
it to a quantum mechanical Hamiltonian, which is required if one is to build the apparatus out  
of the bricks that are available.   
 
Wigner and others define a pointer variable as some sort of quantum observable which is ``coarse'' in the naive sense of not 
varying much over a large  
subspace of a Hilbert space. This famously fails to convert quantum superpositions into classical  
probabilities.  
 
The theory of the Ming Effect uses a novel type of pointer variable: an amplifying apparatus  
is one of the negative temperature systems such as 
$\Cal H_n$.  A measurement apparatus is the thermo-dynamic limit of quantum negative temperature amplifiers, and is a 
classical dynamical system $\Omega_\infty.$ A pointer variable for $\Omega_\infty$ is a macroscopic observable $\{f_n\}$ for 
the sequence $\{\Cal H_n\}$ as above, such that the expectation of $f_n$ is coupled to the initial state of the microscopic 
system $\Cal H_0$ and a fixed class of states of $\Cal H_n$ (regarded as a state of readiness to detect).  The main result of 
this note is that pointer variables exist that satisfy further the probabilistic laws of quantum measurement.  (It seems that the 
condition of being a pointer variable puts strong constraints on the $f_n$.) We now define the family $f_n$, which in the limit, 
becomes the pointer position of the measuring apparatus.  There is a basis of $\Cal H_n^{com}$ consisting of separable 
vectors of the form $\psi_\epsilon\otimes\vert i>$ where $\epsilon$ is 0 or 1.  Let $C$ be the set of basis vectors such that all 
but a negligible number of the $d_i$ for $i<n/2$ are 1 and all but a negligible number of the others are 0.  (This is the device 
being `cocked' and ready to detect.  It is very far from being a stable state, in the limit.)  By negligible, we mean that as a 
proportion of $n$, it goes to zero as $n$ increases.  For $w_n$ any norm one state of the combined system, let $c_i$ be the 
Fourier coefficients of $w_n$ with respect to the cocked basis vectors, i.e., those in $C$.  Define 
$f_n(w_n)=1-\sum_i|c_i|^2$.

Now we are interested in phenomena in the limit as $n$ approaches $\infty$, yet one cannot  
directly compare the arguments of $f_n$ with those of $f_{n+1}$.   
Instead of comparing individual values of these functions,  
one compares time or phase averages of the various $f_n$ as $n$ varies, 
in keeping with the procedures of classical Statistical Mechanics.   
(Phase averages would be taken over the submanifold of accessible states, it is more convenient 
for us to deal with time averages.  Time averages have been made the basis for von Plato's theory[25]  
of the meaning of probability statements.) 
For any abelian dynamical variable $f$, define $<f>=\lim_{T\rightarrow\infty} {1\over T} 
\int_0^T f(e^{\frac{2\pi}h  tH_n}\cdot v)dt$ possibly depending on $v$.   

\centerline {Why time averages?}
 
\vskip -.12in
The Gibbs program usually starts here by postulating a canonical distribution of some 
kind and considering phase averages.  Ford--Kac--Mazur do the same.  
Work of Mazur and Kim following on the above has shown that in the thermodynamic limit, 
the result is largely independent of the initial probability distribution postulated.
The heat bath need not even  be in equilibrium.
It is widely felt, therefore, that something deeper is going on here.  Wiener always 
used time averages to define the auto-correlation function, since for ergodic stationary 
stochastic processes these agree with phase averages.  If the underlying deterministic 
dynamics were ergodic, the time averages would be almost independent of $v$ and 
equal to the phase averages.  Khintchine has shown that for a certain class of abelian 
dynamical variables, even if the dynamics is not ergodic, the phase averages in the 
limit will be equal to the limit of the time averages for those particular dynamical 
variables.  Since we are passing to the thermodynamic limit anyway, the results of 
Mazur, Kim, and others support Khintchine's insight, although the exact range of 
validity of this principle remains unknown.

It is standard practice in theoretical physics to prefer time averages to phase 
averages for probabilities, and so if we take as our goal, to derive the probabilities 
of quantum mechanics from the first three axioms, we need to calculate time averages.
The standard textbook of Landau--Lifschitz for example says that probabilities are 
infinite time averages, and the probability that a system will be found in the 
region $M$ of phase space is, by definition, 
$$\lim_{T\rightarrow\infty} \frac1T \int _0^T \chi_M(v(t)) dt,$$
where $\chi_M$ is the characteristic function of the measureable set $M$.
(This can be criticised from a foundational point of view, but it should only be replaced by a definition that 
yields the same answers because the standard procedure is well-supported experimentally
\dots) and therefore we study the 
thermodynamic limits of time averages.

The reason Wiener always studied time-averages was because, as he and, after him, 
Gelfand, pointed out, the idea of measurement in classical Statistical Mechanics 
is modelled by a time average, since the characteristic relaxation time 
of the measuring device, which is macroscopic compared to the individual degrees 
of freedom, is almost infinite by comparison with the time scale of the underlying 
deterministic dynamics.  von Plato traces this idea to Einstein, and 
von Plato's ergodic theory of the meaning of probability 
makes the probabilistic expectation of the value of $f$ 
equal \it by definition \rm to the time average of $f$.  

There is no use trying to devise a probability distribution for the states of 
the amplifying apparatus as a whole because we control quite a few parameters.
In such a situation it is standard[17], p.\ 50f., procedure to introduce the submanifold of 
``accessible states.''
We know that the amplifying apparatus is in a particular sort of state, a state 
of readiness to detect.  Only a small subspace, $C$, of the total phase space is accessible 
to the apparatus, then.  It is easy to see that our results are completely 
independent of which vector in $C$ we start at.  
The question of what probability distribution on the states of the amplifier could 
be derived by analogy to Khintchine's derivation of the canonical distribution,
but carried out for negative temperatures instead of equilibrium, is therefore 
irrelevant to the question at hand, although of sufficient interest in its own 
right.  It would, however, remove us from the only published coherent theory of 
the meaning of probability assertions that is closely tied to the frequency theory, 
von Plato's ergodic theory of probability.
 
Let the incident particle be in the state described by any (normalised) wave function in $\bold C^2$.  Let it be 
$v_0=a_0\psi_0+a_1\psi_1$.  The amplifier is in the state $|111\dots 000>$ in $C$.   
 
Consider $f_\infty=\lim_{n\rightarrow \infty} <f_n>$. 
Consider a typical trajectory in the manifold  
of accessible states inside of $\Cal H_n^{com}$ (i.e., states which the dynamical system can reach starting from a state in 
$C$). 
It is elementary to calculate  
$\lim_{n\rightarrow\infty} <f_n>$, it is $|a_1|^2$. 
(Nothing is altered if we suppose the amplifier is in a mixed state given by some  
probability distribution supported on the span of $C$.) 

\centerline{Why define this $f_n$?}

     Up until now it has been thought impossible for a pointer variable 
to exist for any model of this type that will agree with the probabilisitic 
axioms.  We are about to succeed in showing that this is not true, it is possible.
How unique is it, and what other macroscopic variables will *not* verify those 
axioms is a question which is too large to answer yet.  To be a macroscopic 
variable at all is a strong constraint, it means the limit of the sequence
$$g(v_o,v_1,v_2,\dots)=\lim_{n\rightarrow\infty} f_n(v_o\otimes v_1\otimes 
v_2\dots\otimes v_n)$$
is a tail event in the theory of probability, and there seems to be some kind 
of zero-one law which governs these.  But tail events are highly non-unique, 
and even if we add the constraint of being a pointer variable in the sense 
that the macroscopic variable must be correlated with the microscopic particle's 
incoming sharp states, it is still non-unique.  We could, for example, take 
an indicator function on the manifold $C$ and get a pointer variable that 
would disobey the probability axioms.  

There are  intuitive reasons for selecting the sequence of $f_n$ which we explicitly 
constructed. 
The first one, which rules out the indicator function, is that the phase functions 
in Classical Statistical Mechanics with physical significance are supported on 
sets of positive measure, not zero or negligible measure.
The second one is that in the debate, which exists, about whether the wave function 
is physical or not, one notices that although the wave function may be rather 
difficult to measure completely with only one measurement, at least one can 
get a physical grip on the $|c_i|^2$'s, they can be measured by repeating 
experiments since they are, after all, operationalised as frequencies.  So a 
physically significant phase function could be constructed out of those 
amplitudes.  Thirdly, the classical phase functions which have thermodynamic 
significance are sum functions, sums of identical functions of each degree of 
freedom, so it is natural to study something like the $f_n$.

\centerline{Continuing}

We will next find a classical dynamical system $\Omega_\infty$ which has a mixed  
state $X$, which depends on $v_0$, and a classical dynamical variable $F$ whose  
expectation values match these limits.  
 
We search for $\Omega_\infty$, $F$, and $X_{v_0}$ as above, satisfying  
$$\int_{\Omega_\infty}FdX_{v_0}=\lim_{n\rightarrow\infty}<f_n>.$$ 
Let the state of the (classical limit) measurement apparatus where the pointer position  
points to cocked (and hence, an absence of detection) be the point $P_0$.  Let the Ming state  
where the excited states of the apparatus are proceeding from out of its initial cocked  
state, and travelling steadily towards the right, be $P_1$. 
Then $\Omega_{\infty}=\{P_0,P_1\}$.  The dynamical variables on this space are generated by 
the characteristic functions of the two points, $\chi_{P_0}, \chi_{P_1}$.  
Let $F$ be $\chi_{P_1}$. It is the pointer position which registers detection. 
The mixed state of  $\Omega_{\infty}=\{P_0,P_1\}$ which gives the right answer    
when the incident particle is in state $v_0$ is the probability distribution which gives  
$P_0$ the weight $|a_0|^2$, and $P_1$ the weight $|a_1|^2$.   
This is precisely what it means to say the the measuring apparatus will register the  
presence of the particle with probability $|a_1|^2$, and fail to register with probability  
$|a_0|^2$.   
 
As the discussion of foundational matters in Khintchine[17] makes clear, thermodynamic  
limits do not really exist, and they need not obey the laws of physics.  They are  
merely convenient devices for organising our thoughts about calculating clever  
approximations to answering questions about finite systems with a large but fixed  
number of degrees of freedom.  Since many approximation techniques for large $n$  
are asymptotic expressions which diverge, it is merely a technical convenience to  
re-normalise or re-scale the question with $n$ to introduce convergence.  There  
was no physical significance to the divergence since $n$ is not in fact a variable  
but a parameter.   
     We renormalised Planck's constant $h$ to be equal to zero in the limit,  
by letting it be proportional to $1\over n$.   

 
     Of course there are many properties of $\Cal H_{1.1\times10^{23}}$ which can  
not be well-studied by neglecting $h$.  But we have just proved that $<f_n>$ 
is not one of them.  Thus, although there are many topics in the physics of  
amplifying apparatus which would be poorly served by taking the limit as  $h\rightarrow 0$,  
the pointer variable is not one of them.  Bohr had this idea, in words.  He said that the  
pointer position was classically describable, meaning that we should be able to study it as  
part of the physical description of the apparatus which does not vary appreciably if  
we neglect Planck's constant. 
The method introduced by this note is the only way yet known to make this precise.  It,  
by coincidence, agrees perfectly with the method of Khintchine and others following the  
Gibbs program.

\centerline{\bf Bibliography}

\noindent [1] H.\ Araki and M.\ Yanase,  Phys.\ Rev.\ 120 (1960), 622. 

\noindent [2] J.\ Bell, Helv.\ Phys.\ Acta 48 (1975) 447, Phys. World 3 (1990), 33.

\noindent [3] F.\ Benatti, \it Deterministic Chaos in Infinite Quantum Systems\rm, Trieste, 1993. 

\noindent [4] W.\ Burnside,  ``On the Idea of Frequency,'' Proc.\ Camb.\ Phil.\ Soc.\rm, \bf 22 \rm (1925), 726.

\noindent [5] M.\ Collet, G.\ Milburn and D.\ Walls, Phys.\ Rev.\ D 32 (1985), 3208. 

\noindent [6] A.\ Daneri, A.\  Loinger and M.\ Prosperi, Nuclear Phys.\ 33 (1962), 297.   

\noindent [7] C.\ Darwin and R.\ Fowler, Phil.\ Mag.\rm, \bf 44 \rm (1922), 450, 823.

\noindent [8] E.\ Farhi, J.\ Goldstone and S.\ Gutmann, Ann.\ Phys.\ 192 (1989), 368.

\noindent [9] G.\ Ford, M.\ Kac and P.\ Mazur, J.\ Math.\ Phys.\ 6 (1965), 504.

\noindent [10] C.\ Gardiner and P.\ Zoller, Quantum Noise, pp.\ 212-229, Berlin, 2000.

\noindent [11] G.\ Ghirardi, A.\ Rimini and T.\ Weber,
\ Phys. Rev. D 34 (1986), 470.

\noindent [12] H.\ Green, Nuovo Cimento 9 (1958), 880.  

\noindent [13] K.\ Hannabuss, Helv.\ Phys.\ Acta 57 (1984), 610, Ann.\ Phys.\ 239 (1995), 296.

\noindent [14] K.\ Hepp, Helv.\ Phys.\  Acta, 45 (1972), 237.

\noindent [15] J.\ Jauch, E.\ Wigner and M.\ Yanase, Nuovo Cimento 48 (1967), 144.

\noindent [16] J.\ Johnson, in \it Quantum Theory and Symmetries III\rm, Cincinnati, 2003, ed.\ by Argyres \it et al\rm., Singapore, 2004, 133.

\noindent [17] A.\ Khintchine, \bf Statistical Mechanics\rm , New York, 1949, p.\ 11f.  

\noindent [18] S.\ Kim, J.\ Math.\ Phys., \bf 15 \rm(1974), 578. 

\noindent [19] A.\ Kolmogoroff, in \it Mathematics its Content, Methods, and Meaning\rm, ed.\ by Aleksandroff, Kolmogoroff, and Lavrentieff, Moscow, 1956 p.\ 239. 

\noindent [20] L.\ Landau and E.\ Lifshitz, \it Statistical Physics\rm, London 1958, p.\ 3.

\noindent [21] J.\ Lewis and H.\ Maassen, Lecture Notes in Mathematics 1055, 245-276, Berlin, 1984.

\noindent [22] J.\ Littlewood, \it A Mathematician's Miscellany\rm, London, 1953, p.\  55f.

\noindent [23] P.\ Mazur, 
in Statistical Mechanics of Equilibrium and Non-Equilibrium, Aachen, 1964, ed.\ by Meixner, Amsterdam, 1965, 69.

\noindent [24] R.\ Minlos, {\it Introduction to Mathematical Statistical Physics}, Providence, 2000, p.\ 22.

\noindent [25] J.\ von Plato, Ergodic Theory and the Foundations of Probability, in 
Brian Skyrms and William Harper (eds.), Causation, Chance, and Credence, Proceedings of the Irvine Conference on Probability and Causation, vol.\ 1, pp.\ 257-277, Kluwer, 1988.

\noindent [26] J.\ Schwinger, J.\ Math.\ Phys.\  2 (1961), 407.

\noindent [27] E.\ Wigner, Am.\ Jour.\ Phys.\ 31 (1963), 6.

\noindent [28] H.\ Zeh, in \it Proceedings of the II International Wigner Symposium\rm, Goslar, Germany, 1991, ed.\ by  Doebner, Scherer and Schroeck, Singapore, 1993, 205.

\noindent [29] W.\ Zurek, Phys.\  Rev.\  D 26 (1982), 1862.

\end